\documentclass{osa-article}

\journal{oe}

\usepackage{graphicx}
\graphicspath{{figures/}}

\begin{document}
% ******************************************************************************

\title{Generation of intense terahertz radiation by two-color filamentation of
	   laser pulses with different wavelengths}

\author{
	Vladimir Yu. Fedorov\authormark{1,2,*} and
	Stelios~Tzortzakis\authormark{1,3,4,**}
}

\address{
	\authormark{1}Science Program, Texas A\&M University at Qatar, P.O. Box
				  23874, Doha, Qatar\\
	\authormark{2}P.~N.~Lebedev Physical Institute of the Russian Academy of
				  Sciences, 53 Leninskiy Prospekt, 119991, Moscow, Russia\\
	\authormark{3}Institute of Electronic Structure and Laser (IESL), Foundation
				  for Research and Technology - Hellas (FORTH), P.O. Box 1527,
				  GR-71110 Heraklion, Greece\\
	\authormark{4}Materials Science and Technology Department, University of
				  Crete, 71003, Heraklion, Greece
}

\email{
	\authormark{*}v.y.fedorov@gmail.com\\
	\authormark{**}stzortz@iesl.forth.gr
}

\begin{abstract}
	We theoretically study the generation of terahertz (THz) radiation by
	two-color filamentation of ultrashort laser pulses with different
	wavelengths.
	We consider wavelengths in the range from 0.6 to 10.6~$\mu$m, thus covering
	the whole range of existing and future powerful laser sources in the near,
	mid and far-infrared.
	We show how different parameters of two-color filaments and generated THz
 	pulses depend on the laser wavelength.
	We demonstrate that there is an optimal laser wavelength for two-color
	filamentation that provides the highest THz conversion efficiency and
	results in generation of extremely intense single cycle THz fields.
\end{abstract}

% ******************************************************************************
\section{Introduction}
% ******************************************************************************
In the electromagnetic spectrum, the terahertz (THz) frequency range is located
in-between the microwave and far-infrared frequencies.
For many reasons THz radiation attracts a lot of attention~\cite{Zhang2017,
Tonouchi2007}.
First, most molecules (especially large biological molecules) have
characteristic structural resonances at THz frequencies.
Then, the THz part of the electromagnetic spectrum is located at the boundary
between spectral ranges characteristic for high frequency electronics and
photonics.
Therefore, THz devices could play a role of a bridge between these two
technologies.
Another advantage of THz radiation is its high penetration depth in many
materials like plastics, wood, paper, clothings, etc.
What is more important, unlike x-rays, THz photons have very low energy and do
not ionize matter.
As a result, one can use THz radiation in a large number of applications in
industrial quality control, medical diagnostics and treatment, homeland security
and many others.

Despite the high potential of THz radiation for applications, there is still a
huge lack of powerful THz sources.
Nowadays, the most powerful table-top sources of THz radiation are based on
optical rectification in nonlinear crystals or two-color
filamentation~\cite{Lewis2014,Reimann2007}.
Using optical rectification one can reach 3.7\% of THz conversion efficiency
(ratio of generated THz energy to energy of input laser pulse) and up to 0.9~mJ
of THz pulse energy~\cite{Huang2014,Vicario2014,Fulop2014,Shalaby2015}.
However the spectrum of the THz pulses generated by optical rectification is
quite narrow and limited to frequencies below 5~THz.
In turn, with two-color filamentation a typical THz conversion efficiency is
lower ($\sim$0.01\%) and the generated THz pulses are less energetic (up to
30~$\mu$J in gases and up to~80 $\mu$J in liquids)~\cite{Kim2008,Oh2014,Kuk2016,
Dey2017}.
Nevertheless, compared to optical rectification, THz pulses generated by
two-color filamentation are considerably shorter and have much broader spectra
(>50~THz).
Therefore, in spite of the lower energy, the peak power of such THz pulses can
be much higher.
Thus, the two-color filamentation technique is a promising tool for generation
of THz pulses, powerful enough for studies of nonlinear interactions of THz
radiation with matter.
Additionally, with two-color filamentation one can generate THz radiation at
remote distances, close to the object of interest, which allows one to overcome
strong diffraction of THz fields and their high absorption in atmospheric water
vapor~\cite{Wang2010,Wang2011,Daigle2012,Liu2016}.

Until recently, the majority of experiments on THz generation by two-color
filamentation were conducted with Ti:Sapphire laser systems operating at a
central wavelength of 0.8~$\mu$m.
Nevertheless, an experimental study of two-color filamentation at longer
wavelengths showed more than tenfold growth of the THz conversion efficiency
when the wavelength of the fundamental laser pulse was increased up to
1.8~$\mu$m, though beyond this wavelength the efficiency dropped down
again~\cite{Clerici2013}.
The evidence of stronger THz generation at mid-infrared wavelengths was also
obtained from Particle in Cell (PIC) simulations of 4~$\mu$m single color laser
pulses interacting with gas targets (though without consideration of nonlinear
propagation effects)~\cite{Wang2011OL}.

Recently appeared powerful laser sources, operating at a central wavelength of
3.9~$\mu$m~\cite{Mitrofanov2015}, have opened the way to experimental studies of
THz generation by laser pulses at the mid-infrared spectral region.
In a recent theoretical study, we showed that two-color filamentation of
mid-infrared 3.9~$\mu$m laser pulses allows one to generate single cycle THz
pulses with multi-millijoule energies and record conversion efficiencies
reaching 7\% (more than two orders of magnitude higher than for 0.8~$\mu$m laser
pulses)~\cite{Fedorov2017,Fedorov2018}.
Later, similar results were obtained in~\cite{Nguyen2018}.
Our recent experiments on two-color filamentation of 3.9~$\mu$m laser
pulses~\cite{Koulouklidis2018} confirmed the theoretical predictions.
In our theoretical predictions, the focused single cycle THz pulses can reach
peak electric and magnetic fields at the GV/cm and kT level respectively.
These high conversion efficiencies and field strengths are the result of several
factors: stronger photocurrents due to larger ponderomotive forces, negligible
walk-off between the fundamental and second harmonic, longer and wider plasma
channels, additional field symmetry breaking by generated high
harmonics~\cite{Fedorov2018}.

Despite the above studies there is still an open question about the optimal
laser source in terms of its central wavelength for THz generation by two-color
filamentation.
To answer this question, in this paper we use numerical simulations to study the
generation of THz radiation by two-color filamentation of laser pulses with
different wavelengths.
Our theoretical modeling and simulations have been extensively and successfully
benchmarked with experiments at both 0.8~$\mu$m and at 3.9~$\mu$m wavelengths.
Here we consider central laser wavelengths in the range from 0.6 to 10.6~$\mu$m,
thus covering the whole range of existing and future high power laser sources.
We show how the parameters of two-color filaments and generated THz pulses
depend on the laser wavelength.
We demonstrate that for two-color filamentation there is an optimal wavelength
of the laser source that provides the highest THz conversion efficiency.

% ******************************************************************************
\section{Model}
% ******************************************************************************
A detailed description of the model that we use in our simulations can
be found in~\cite{Fedorov2018}.
Briefly, for the simulations of two-color filamentation we use the
unidirectional pulse propagation equation (UPPE) coupled with a rate equation
for plasma density~\cite{Kolesik2002,Kolesik2004,Couairon2011}.
The UPPE equation is free of slow envelope and paraxial approximations, and has
been successfully applied in previous studies of THz generation by near-infrared
two-color filaments by many groups beyond ours~\cite{Babushkin2010,Berge2013,
Andreeva2016}.
To precisely describe the dispersion of air refractive index we use the data on
absorption lines of oxygen, nitrogen, and carbon dioxide form the HITRAN
database~\cite{HITRAN}.
In our simulations we assume that air is dry, thus neglecting the absorption of
the generated THz radiation by atmospheric water vapor.
Such assumption is justified since we consider only focused laser pulses
propagating over short distances.
Under these conditions, the whole experimental setup can be placed into a purge
gas chamber filled with dry air, as we usually do in our experiments.
To calculate the field ionization rate we use the wavelength-dependent
Perelomov-Popov-Terent'ev formula~\cite{Perelomov1967}.

As initial condition for the UPPE equation we use the following two-color field:
\begin{equation}
    E = \exp\left(-\frac{r^2}{2a_0^2} - \frac{t^2}{2\tau_0^2}\right)
        \left[E_1 \cos\left(\omega_0t\right)
            + E_2 \cos\left(2\omega_0t\right)\right],
\end{equation}
where $r^2=x^2+y^2$, $a_0=4/2\sqrt{\ln{2}}$~mm is the beam size (4~mm FWHM),
$\tau_0=100/2\sqrt{\ln{2}}$~fs is the pulse duration (100~fs FWHM), $\omega_0$
is the central frequency of the fundamental pulse, while $E_1$ and $E_2$ being
the amplitudes of the fundamental and the second harmonic pulses, respectively.
The initial pulse is focused by a lens with the focal distance $z_f$=200~mm.
To simulate the focusing we multiply each Fourier harmonic of the field $E$ by
a factor $\exp\left[-i(\omega/c_0)r^2/(2z_f)\right]$, where $\omega$ is the
frequency of the corresponding harmonic and $c_0$ is the speed of light in
vacuum.

In our simulations we consider the fundamental pulses with central wavelength
$\lambda_0=2\pi c_0/\omega_0$ ranging from 0.6 to 10.6~$\mu$m.
The energy $W$ for each wavelength was chosen in such a way that the peak power
$P$ of the corresponding single-color pulse at wavelength $\lambda_0$ is equal
to 1.2$P_\text{cr}$, where $P_\text{cr}$ is the critical power of self-focusing
in air at this wavelength.
Therefore, since $P_\text{cr}\propto\lambda_0^2$, the input energy $W$ also
growths quadratically with the wavelength, starting from 0.69~mJ for
$\lambda_0$=0.6~$\mu$m and reaching 216~mJ for $\lambda_0$=10.6~$\mu$m.
For all wavelengths in our studies the fundamental and second harmonic pulses
hold, respectively, 95\% and 5\% of the input energy $W$.

The fixed ratio of peak power to critical power is a fundamental condition for
comparing filamentation of laser pulses with different wavelengths.
If one would fix the input energy $W$ of the laser pulses the comparison would
be impossible, since at shorter wavelengths one would be in the regime of
nonlinear propagation while for far-infrared wavelengths the propagation would
be purely linear.

The whole propagation distance in our simulations is equal to 400~mm, which is
much shorter than in most previous simulations of mid and far-infrared
filamentation, where the propagation distances were tens and hundreds of
meters~\cite{Geints2014,Panagiotopoulos2015,Panov2016}.
Contrary to filamentation of long-wavelength laser pulses at long distances,
where the intensity clamping happens due to shock wave formation rather than by
defocusing in plasma~\cite{Panagiotopoulos2015}, for focused long-wavelength
laser pulses at short distances plasma becomes the main mechanism of intensity
saturation~\cite{Shumakova2018}.

% ******************************************************************************
\section{Results}
\begin{figure}[htbp]
	\centering\includegraphics{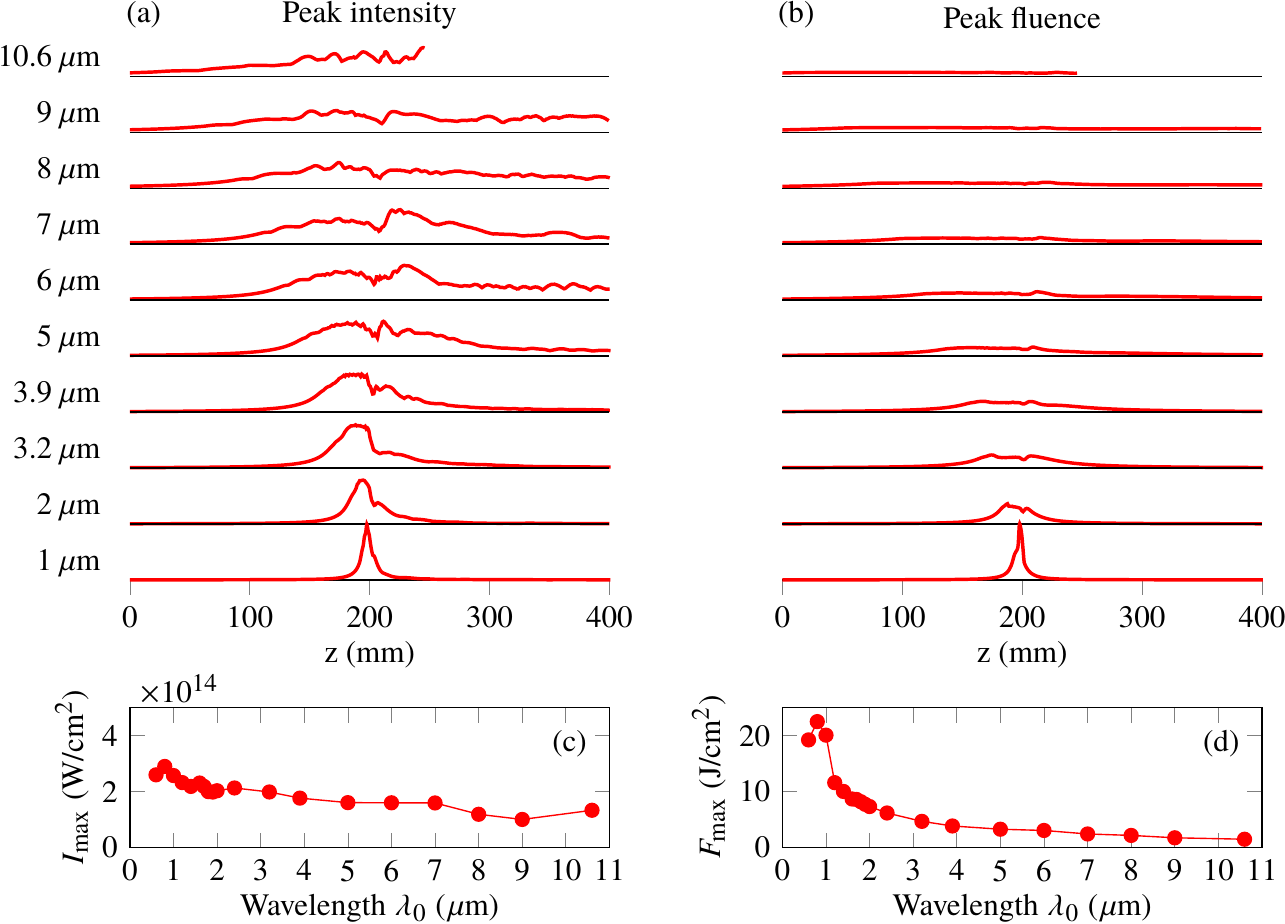}
	\caption{\label{fig:intensity}%
		(a) Normalized peak intensity and (b) peak fluence versus propagation
		distance for several wavelengths $\lambda_0$ of the fundamental pulse.
		The normalization factor is the same for all wavelengths.
		(c) Maximum peak intensity $I_\text{max}$ and (d) maximum peak fluence
		$F_\text{max}$ versus the wavelength $\lambda_0$ of the fundamental
		pulse.}
\end{figure}

% ******************************************************************************
In this section we present the results of our simulations of two-color
filamentation at different laser wavelengths.
In Fig.~\ref{fig:intensity} we show how the peak intensity and peak fluence
(integral of the laser pulse intensity over time) vary over distance for laser
pulses with different wavelength $\lambda_0$.
One can see that the maximum peak intensity $I_\text{max}$ is almost independent
of the pulse wavelength (see Figs.~\ref{fig:intensity}(a) and (c)).
However, for longer wavelengths high intensities exist over longer distances
(see Fig.~\ref{fig:intensity}(a)) which means that laser pulses with longer
wavelengths produce longer filaments.
In Figs.~\ref{fig:intensity}(b) and (d) we see that the maximum value of peak
fluence $F_\text{max}$ decreases with the increase of the pulse wavelength.
As we will see in the following, such behavior of the peak fluence can be
explained by the wider supercontinuum spectrum for laser pulses with longer
wavelength and, as a result, more effective pulse shortening during
filamentation (we remind that fluence is the integral of the laser pulse
intensity over time).

\begin{figure}[htbp]
	\centering\includegraphics{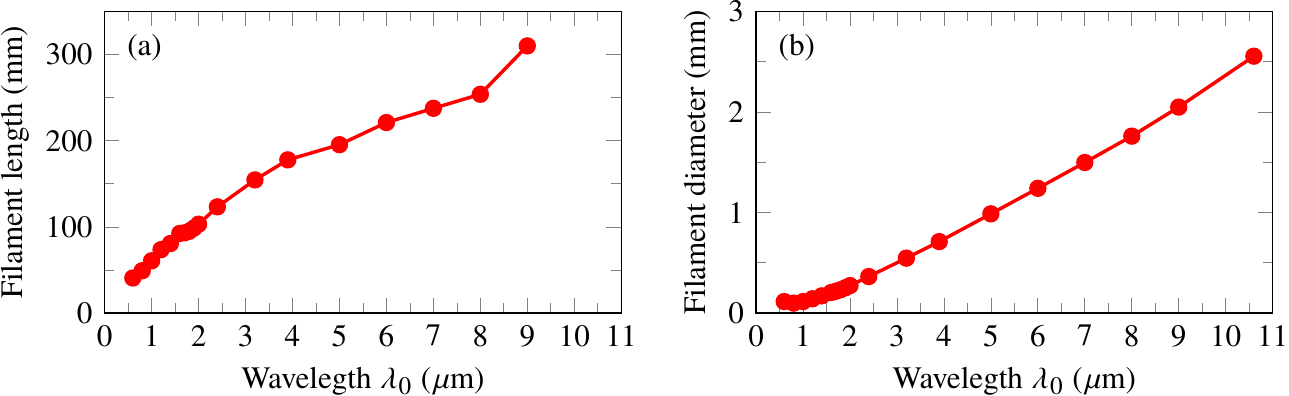}
	\caption{\label{fig:radius}%
		(a) Filament length (FWHM) and (b) filament diameter (FWHM) versus the
		wavelength $\lambda_0$ of the fundamental pulse.}
\end{figure}

Figure~\ref{fig:radius} shows the dependence of the filament length (calculated
as the FWHM length of the fluence distribution along $z$) and filament diameter
(calculated as the minimal along $z$ FWHM diameter of the fluence distribution)
on the wavelength $\lambda_0$ of the fundamental pulse.
We see that both the length and diameter of the filament increase as a function
of the pump wavelength.
Thus, for laser pulses with longer wavelength the filaments are not only longer
but also thicker.
As expected, the same observation is true for the corresponding plasma channels:
they also become longer and thicker with increase of $\lambda_0$.

\begin{figure}[htbp]
	\centering\includegraphics{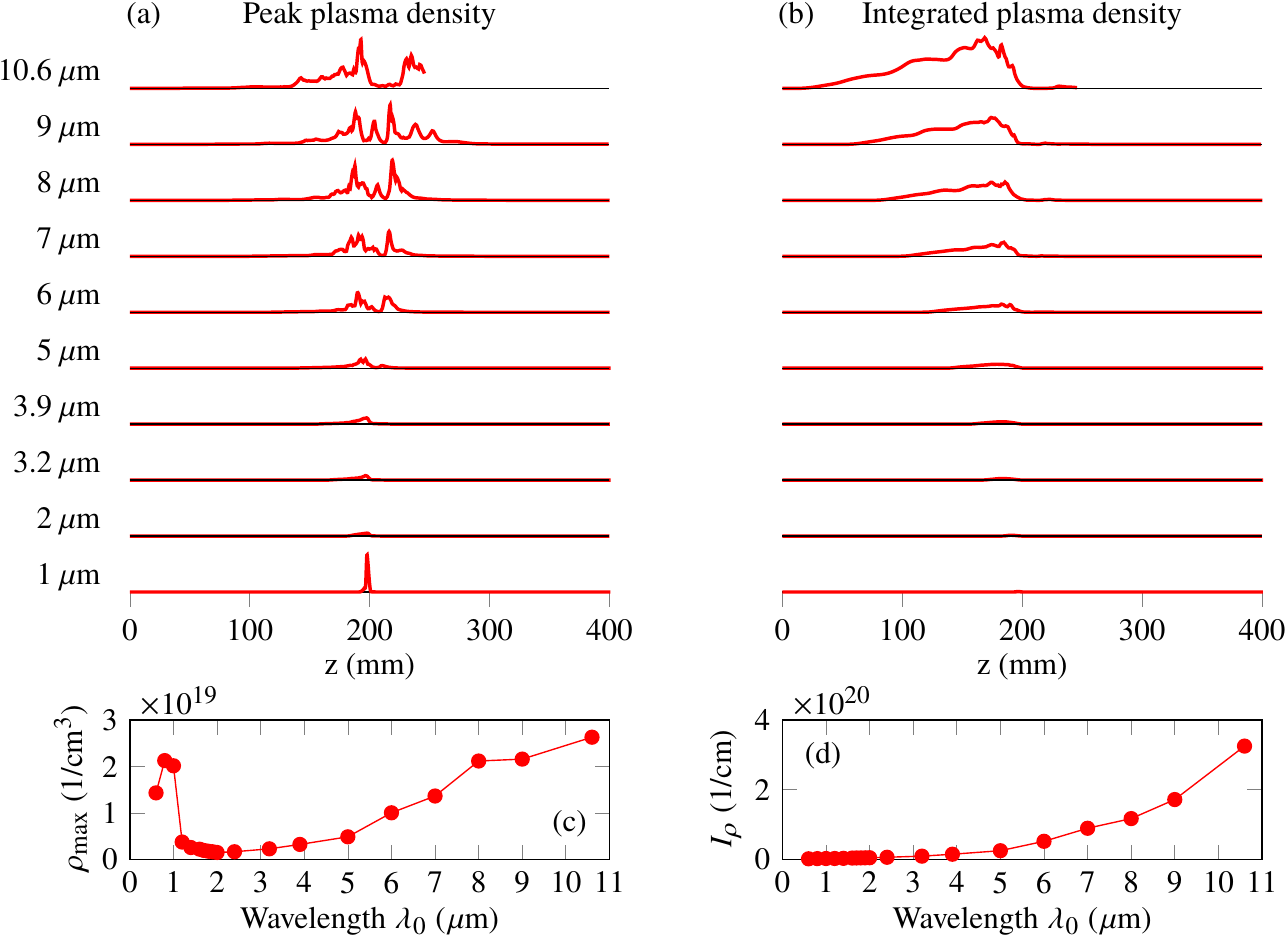}
	\caption{\label{fig:plasma}%
		Normalized peak (a) and integrated over radius (b) plasma densities
		versus propagation distance for several wavelengths $\lambda_0$ of the
		fundamental pulse.
		The normalization factor is the same for all wavelengths.
		(c) Maximum peak plasma density $\rho_\text{max}$ and (d) maximum
		integrated plasma density $I_\rho$ versus the wavelength $\lambda_0$ of
		the fundamental pulse.}
\end{figure}

Figure~\ref{fig:plasma} shows the dependence of peak and integrated (over
radius) plasma densities on propagation distance for laser pulses with different
wavelength $\lambda_0$.
In Figs.~\ref{fig:plasma}(a) and (c) we see that the maximum peak plasma density
$\rho_\text{max}$, being high for near infrared pulses, starting from
approximately $\lambda_0$=1~$\mu$m, drops down.
Then, for mid and far-infrared pulses, $\rho_\text{max}$ increases again and
reaches similar values to the ones for near-infrared pulses.
This transition can be explained by the interplay between the field and
avalanche ionizations, since the avalanche ionization becomes more important at
longer wavelengths.
For near infrared and shorter wavelengths most of the plasma free electrons are
produced by the field ionization.
With increase of the wavelength, the probability of the field ionization drops
down and we start to see less plasma density.
However, the ponderomotive forces exerted by mid and far-infrared laser pulses
become strong enough to effectively accelerate free electrons, which results
in enhanced avalanche ionization and generation of more plasma.
This behavior is confirmed by our simulations, where we see almost no plasma
generation for mid and far-infrared pulses when we switch-off the avalanche
ionization.

Figure~\ref{fig:plasma}(d) shows that the maximum value of the integrated plasma
density, $I_\rho$, monotonically increases with increase of the pulse central
wavelength $\lambda_0$.
Taking into account that the peak plasma densities for far and near infrared
pulses are close (see Fig.~\ref{fig:plasma}(c)), we conclude that this increase
of $I_\rho$ happens because plasma channels, produced by laser pulses with
higher $\lambda_0$, are wider (see Fig.~\ref{fig:radius}).
From Fig.~\ref{fig:plasma}(b) we can also conclude that for higher $\lambda_0$
the plasma channels are longer.
For example, for $\lambda_0$=10.6~$\mu$m the plasma channel is being formed
almost immediately after the laser pulse starts to propagate in the atmosphere.
The ponderomotive forces for such long-wavelength laser pulses are so strong
that even a small seed of free electrons produced by field ionization is enough
to generate a huge amount of secondary free electrons through avalanche
ionization.
To sum up, due to longer and wider plasma channels, the total amount of free
electrons produced by laser pulses with longer wavelength increases.

\begin{figure}[htbp]
	\centering\includegraphics{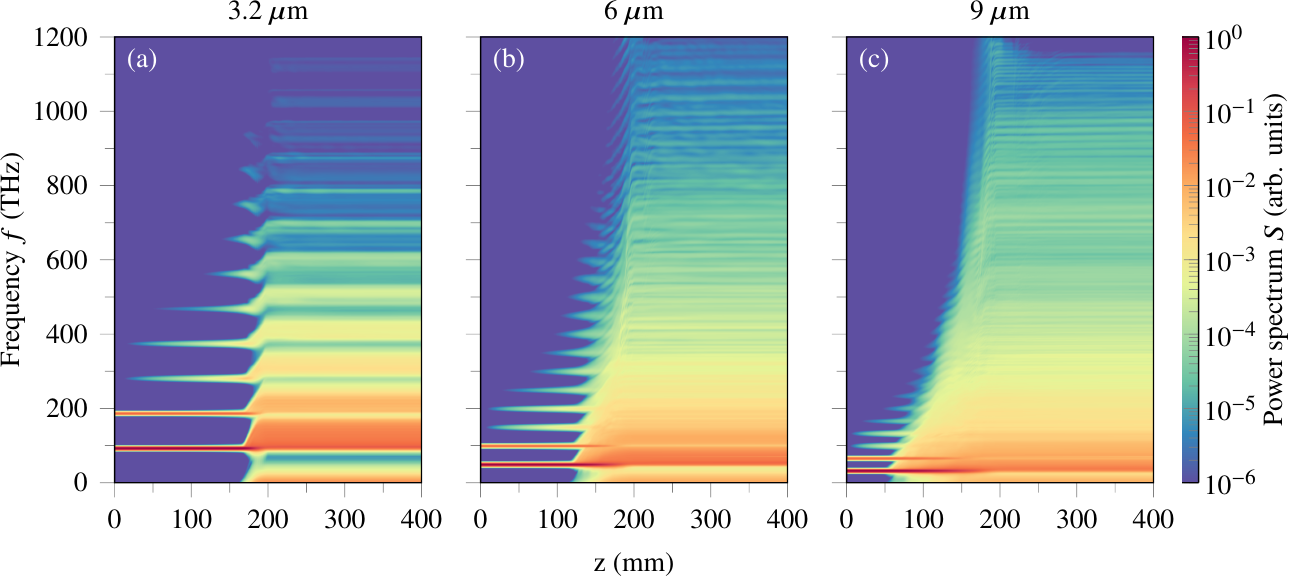}
	\caption{\label{fig:iSzf}%
		Dependence of the integrated over radius pulse power spectrum $S$ on
		the propagation distance $z$ and frequency $f$ for (a) 3.2~$\mu$m, (b)
		6~$\mu$m, and (c) 9~$\mu$m two-color laser pulses.}
\end{figure}

In Fig.~\ref{fig:iSzf} we show the dependence of the integrated (over radius)
power spectrum on the propagation distance $z$ and frequency $f$ for laser
pulses with central wavelength 3.2, 6, and 9~$\mu$m.
For all three wavelengths $\lambda_0$ we see the formation of multiple higher
harmonics.
However, while for $\lambda_0$=3.2~$\mu$m all higher harmonics remain spectrally
separated over the whole propagation distance, for $\lambda_0$=6 and 9~$\mu$m,
after plasma formation, the higher harmonics merge together and form
uninterrupted supercontinua, extending down to the ultraviolet absorption band
of the atmosphere.
This merging of higher harmonics happens, because the distance between the
neighboring harmonics is inversely proportional to $\lambda_0$.
As a result, in case of long-wavelength laser pulses, spectral broadening and
blue shifting in plasma fill the gaps between the harmonics much easier.
As we have shown is our previous study~\cite{Fedorov2018}, the various harmonic
pairs play a significant role in enhancing the plasma photocurrents and
consequently the THz conversion efficiency.

\begin{figure}[htbp]
	\centering\includegraphics{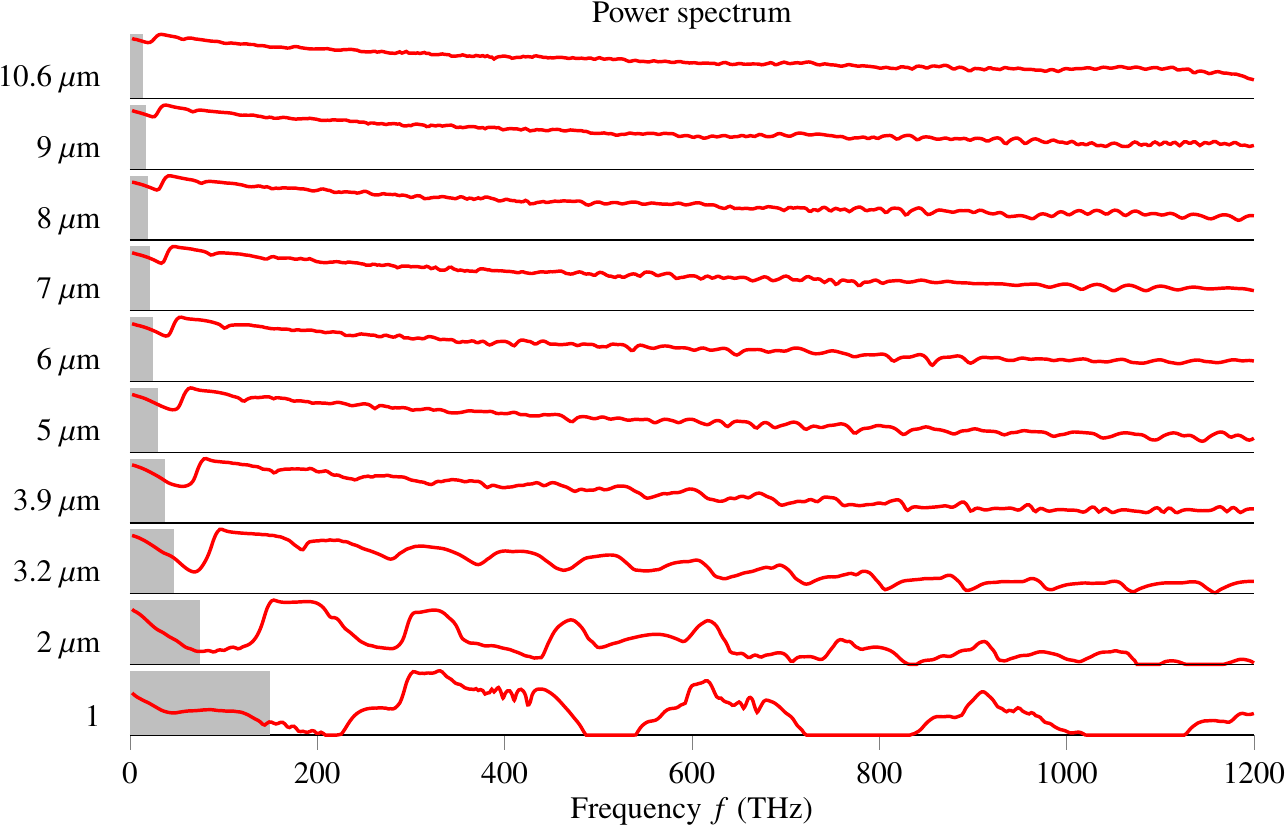}
	\caption{\label{fig:spectrum}%
		Normalized power spectra (integrated over radius) of laser pulses with
		different wavelength $\lambda_0$.
		The plots are in log scale and the normalization factor is the same for
		all wavelengths.
		All spectra are obtained in the point of lens focus at distance 200~mm.}
\end{figure}

Figure~\ref{fig:spectrum} shows the integrated (over radius) power spectra,
obtained at a distance of 200~mm (lens focus), for two-color laser pulses with
different wavelengths $\lambda_0$.
Here again we see that for near and mid-infrared laser pulses we can distinguish
separate higher harmonics generated during nonlinear propagation.
However, for far-infrared laser pulses all harmonics merge together and form
uninterrupted supercontinuum spectra that cover the whole range from the THz up
to ultraviolet frequencies.
This effect also explains the decreasing peak fluence observed in
Fig.~\ref{fig:intensity}(d).

In Fig.~\ref{fig:spectrum} we also see that for all wavelengths $\lambda_0$ a
considerable part of the laser pulse energy transfers to THz frequencies.
However, it is not so straightforward to separate the generated THz radiation
from the rest of the spectrum, especially in the case of far-infrared pulses.
For example, for laser pulses with $\lambda_0$=10.6~$\mu$m, the central
frequency $f_0=\omega_0/2\pi$ is equal to 28~THz, so the fundamental pulse
itself can be called a THz pulse.
In order to give a common definition of the generated THz pulse, independently
of the laser pulse wavelength, we define the generated THz radiation as all
frequencies that lie below half of the laser central frequency, that is, below
$f_0/2$.
For each wavelength $\lambda_0$ in Fig.~\ref{fig:spectrum} the gray shaded area
shows the part of the laser pulse spectrum that, according to our definition,
corresponds to the generated THz pulse.
As one can see, the spectrum of the generated THz pulses narrows down with
increase of $\lambda_0$.

In principle, one could define the generated THz radiation as radiation with
frequencies below some fixed frequency, e.g. 15~THz.
However, such definition, being applied for near and mid-infrared pulses, would
exclude all higher THz frequencies that are generated during two-color
filamentation.
In other words, such definition would neglect the fact that two-color
filamentation is a source of ultra-broadband THz radiation.

\begin{figure}[htbp]
	\centering\includegraphics{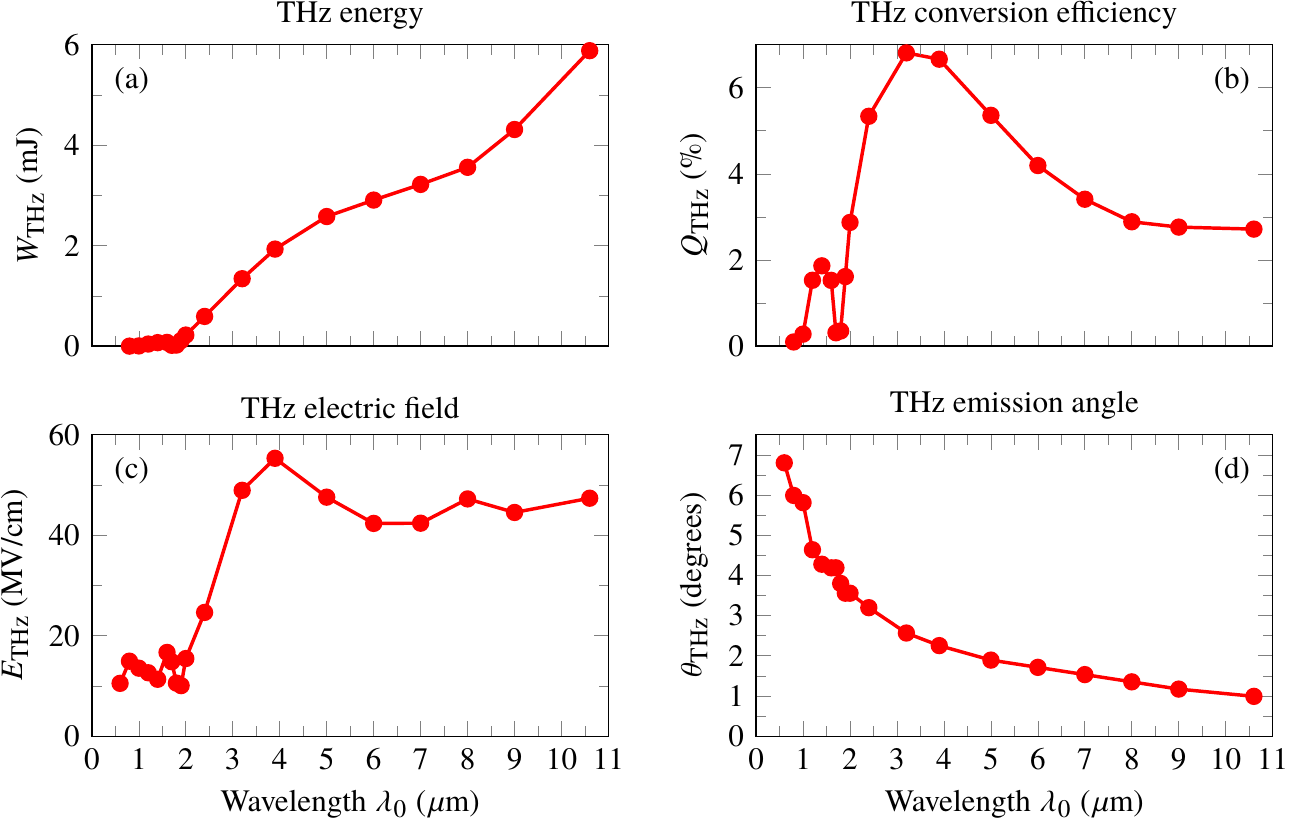}
	\caption{\label{fig:parameters}%
		(a) Energy of generated THz pulse $W_\text{THz}$, (b) THz conversion
		efficiency $Q_\text{THz}$, (c) peak THz electric field $E_\text{THz}$,
 		and (d) THz emission angle $\theta_\text{THz}$ versus the wavelength
 		$\lambda_0$ of the fundamental laser pulse.}
\end{figure}

Figure~\ref{fig:parameters} shows how different parameters of the generated THz
pulses depend on the wavelength $\lambda_0$ of the two-color laser source.
In Fig.~\ref{fig:parameters}(a) we see the dependence of energy of the generated
THz pulses $W_\text{THz}$ on $\lambda_0$.
Since in our simulations we fix the $P/P_\text{cr}$ ratio and
$P_\text{cr}\propto\lambda_0^2$, the input laser energy is also proportional to
the square of $\lambda_0$.
Therefore, in order to have the same or growing THz conversion efficiency while
increasing the laser wavelength, $W_\text{THz}$ should increase at least as
$\lambda_0^2$.
However, according to Fig.~\ref{fig:parameters}(a) the THz energy growths
non-monotonically with wavelength $\lambda_0$.
As a result, the THz conversion efficiency $Q_\text{THz}$ depends on
$\lambda_0$ in a peculiar way (see Fig.~\ref{fig:parameters}(b)):
First, $Q_\text{THz}$ rapidly growths while $\lambda_0$ increases up to
approximately~1.6 $\mu$m, but then drops down, reaching its minimum near
$\lambda_0$=1.8~$\mu$m.
This part of the $Q_\text{THz}$ dependence is very similar to the experimental
observations in~\cite{Clerici2013}.
After 1.8~$\mu$m, THz conversion efficiency starts to growth again and reaches
its maximum value of about 7\% around $\lambda_0$=3.2~$\mu$m.
Then, at longer wavelengths $Q_\text{THz}$ monotonically decreases.
Since the THz conversion efficiency has a global maximum near
$\lambda_0$=3.2~$\mu$m, we can conclude that this laser wavelength is the
optimal one for all THz sources based on two-color filamentation.

Here we should stress the importance of the propagation effects.
According to the photocurrent model~\cite{Kim2007}, that is usually used to
explain the mechanism of THz generation by two-color filamentation, THz energy
should continuously grow with increase of the laser wavelength $\lambda_0$.
Thus, the peculiar dependence of THz energy on $\lambda_0$ is the result of the
complex interplay between linear and nonlinear effects during the laser pulse
propagation.

Figure~\ref{fig:parameters}(c) shows the dependence of the peak THz electric
field $E_\text{THz}$ on the wavelength $\lambda_0$ of the two-color laser
source.
We see that for mid and far-infrared laser pulses $E_\text{THz}$ is higher than
for near-infrared ones.
Though, starting from $\lambda_0\approx3$~$\mu$m the peak THz electric field is
almost independent of $\lambda_0$ and approximately equal to 50~MV/cm.
Note that this field strength corresponds to THz pulses inside the filamentation
zone.
By collecting the THz radiation and refocusing it after the filament, using for
example simple parabolic mirrors, one can reach significantly higher electric
and magnetic THz fields of GV/cm and kT level~\cite{Fedorov2018}.

In the spatial domain the THz radiation generated by two-color filamentation is
emitted into a cone~\cite{Zhong2006,Blank2013,Klarskov2013,Gorodetsky14}.
Figure~\ref{fig:parameters}(d) shows a monotonic decrease of the angle
$\theta_\text{THz}$ of this THz cone as a function of increasing laser
wavelength $\lambda_0$.
This effect is linked to the longer and thicker filament plasma channels
(see Fig.~\ref{fig:radius}) at longer pump wavelengths as explained by the
interference model~\cite{Gorodetsky14}.
Thus, the THz radiation generated by laser pulses with longer wavelengths is
better directed.

% ******************************************************************************
\section{Conclusion}
% ******************************************************************************
In conclusion, we used numerical simulations to study the generation of THz
radiation by two-color filamentation of laser pulses with different wavelength.
We considered laser wavelengths in the range from 0.6 to 10.6~$\mu$m.
This spectral range includes all existing and future high power ultrashort laser
sources.
We have shown that laser pulses with longer wavelengths produce longer and wider
filaments and plasma channels.
The total amount of free electrons produced by laser pulses increases with
increasing laser wavelength.
Also, we demonstrated that two-color filamentation of far-infrared laser pulses
is accompanied by the formation of extremely broad uninterrupted supercontinua
that extend from THz up to ultraviolet frequencies.
We have shown that using mid and far-infrared two-color laser pulses one can
generate THz laser pulses whose electric fields are about 50--60~MV/cm inside
the filamentation zone and can exceed the GV/cm level through further focusing.
Also we demonstrated that the THz radiation for laser pulses with longer
wavelengths is better directed: the angle of the THz conical emission reduces
with increase of the laser wavelength.
Moreover, we have shown that the highest THz conversion efficiency of
$\approx$7\% is reached for two-color laser sources operating at wavelengths
close to 3.2~$\mu$m.
Thus, we found the optimal laser wavelength for THz generation by two-color
filamentation, which opens the way to extreme nonlinear THz science.

% ******************************************************************************
\section*{Funding}
% ******************************************************************************
This work was supported by the National Priorities Research Program grant
No.~NPRP9-329-1-067 from the Qatar National Research Fund (member of The Qatar
Foundation), the H2020 Laserlab-Europe (EC-GA~654148), and the H2020 MIR-BOSE
(EC-GA~737017).

% ******************************************************************************
\bibliography{main}

\end{document}